\documentclass[trackchanges]{aastex701}

\usepackage{amsfonts,amssymb,amsthm,bbm,amsmath}
\usepackage{enumerate}
\usepackage{upgreek}

\usepackage{stackengine}
\usepackage{scalerel}
\usepackage{color,psfrag}
\usepackage{graphicx}
\usepackage{tkz-euclide}
\usepackage{tikz}
\usetikzlibrary{calc}
\usetikzlibrary{decorations.pathmorphing}
\usetikzlibrary{shapes.geometric}
\usetikzlibrary{matrix}
\usetikzlibrary{arrows,decorations.markings}
\usetikzlibrary{shadings,intersections}
\usepackage{caption}
\usepackage{subcaption}

\begin{document}

\title{Distinguish Bardeen-like black holes by Gravitational lensing}

\author[]{Limei Yuan} 
\affiliation{State Key Laboratory of Surface Physics, Center for Astronomy and Astrophysics, Department of Physics, Center for Field Theory and Particle Physics, and Institute for Nanoelectronic devices and Quantum computing, Fudan University, Shanghai 200433, China}
\email[show]{lmyuan24@m.fudan.edu.cn}
\author[gname=Chen-Hung, sname='Hsiao']{Chen-Hung Hsiao}
\affiliation{State Key Laboratory of Surface Physics, Center for Astronomy and Astrophysics, Department of Physics, Center for Field Theory and Particle Physics, and Institute for Nanoelectronic devices and Quantum computing, Fudan University, Shanghai 200433, China}
\email[show]{chsiao@fudan.edu.cn}

\author[]{Yidun Wan}
\affiliation{State Key Laboratory of Surface Physics, Center for Astronomy and Astrophysics, Department of Physics, Center for Field Theory and Particle Physics, and Institute for Nanoelectronic devices and Quantum computing, Fudan University, Shanghai 200433, China}
\affiliation{Hefei National Laboratory, Hefei 230088, China}
\email[show]{ydwan@fudan.edu.cn}

\begin{abstract}
We study Bardeen-like regular black holes without Cauchy horizons via gravitational lensing. In the weak field, the deflection angle receives a positive \(\ell\)-dependent correction, producing a slightly larger Einstein ring. For the galaxy ESO 325‑G004, the predicted ring radius is consistent with current observations. In the strong field, for Sgr A* and M87*, the asymptotic position \(\theta_{\infty}\) remains identical to the Schwarzschild value; however, SDL coefficients are $\ell$-dependent, the angular separation \(s\) increases and the relative flux ratio \(r_{\mathrm{mag}}\) decreases as \(\ell\) increases. Time delays between relativistic images for Sgr A* and M87* also increase mildly with \(\ell\). Our calculated values for these observables remain consistent with current observations. Future strong-field measurements of $\triangle T_{2,1}$, \(s\), and \(r_{\mathrm{mag}}\) may offer a viable test for regular black holes free of Cauchy horizons and may distinguish Bardeen-like from Schwarzschild black holes.
\end{abstract}

\section{Introduction}
Gravitational lensing provides a powerful tool for compact objects and the nature of gravity in both weak and strong fields. Regular black holes resolve the central singularity but may pose an issue with the potential emergence of Cauchy horizons~\cite{Poisson:1990eh, Maeda:2005yd}, such as Hayward and Bardeen black holes \cite{Hayward:2005gi,1968qtr..conf...87B}. The Bardeen-like regular black hole resolves the singularity without causing the Cauchy horizon. An important question arises: can the distinct geometry of this black hole be distinguished from a Schwarzschild black hole through observable lensing effects with modern or future telescopes?
In this work, we systematically study the gravitational lensing signatures of Bardeen-like regular black holes in both weak and strong fields to address the question. We obtain the following key results:

\begin{enumerate}
 
 \item  In the weak field regime, the deflection angle increases with $\ell$, and the theoretical Einstein ring for the Bardeen-like black hole is slightly larger than one in the Schwarzschild case. Using data in Galaxy ESO 325-G004, the calculated Einstein ring is consistent with the observed one within the error bars.
 
 \item  In the Strong Deflection Limit (SDL), deflection angle is computed and SDL observables $(\theta_{\infty},s,r_{\mathrm{mag}})$ are computed using data in Sgr A* and M87*. Obervables $(s,r_{\mathrm{mag}})$ are $\ell$-dependent,
 however, the black-hole shadow radius $\theta_{\infty}$ is $\ell$-independent. 
 \item Time delay $\Delta T_{2,1}$ between the first two images increases with $\ell$ for Sgr A* and M87* cases. The calculated black hole shadows remain consistent with the measured values for M87* and Sgr A*. 
\end{enumerate}

\begin{figure}[h!]
\centering
\begin{tikzpicture}[>=latex,scale=1.0,
    image/.style={very thick,blue},
    source/.style={thick,red,densely dashed},
    axis/.style={gray,dashed}]

    \coordinate (O) at (0,0);       
    \coordinate (L) at (6,0);       
    \coordinate (S) at (11.5,1.02);   
    \coordinate (I) at (5.57,1.76);    
    
    \coordinate (J) at (6.565,2.1);  
    \coordinate (Y) at (6.53,2.12);  

    \draw[axis] (O) -- (12,0) node[below right]{optical axis};
    \draw[axis] (11.5,-1.2) -- (11.5,2.2) node[right]{source plane};

    \draw[thick] (L) circle (0.16);
    \fill (L) circle (1pt) node[below] {$L$};

     \draw[image] (O) .. controls (5.2, 2.2) and (6.3, 2.2) .. (S);

    \draw[source] (O) -- (S);
    \draw[dashed,blue, thick] (L) --node[left] {$\rho_0$} (I);
    \draw[dashed, blue, thick] (S) -- (3, 2.85) node[left, black] {$Y$};
    \draw[dashed, blue, thick] (O) -- (9, 3.72) node[right, black] {$Z$};

    \draw[dashed,blue, thick] (L) -- node[right] {$b$} (Y);  
    \node[left] at (Y) {$X$};

    \node[right=2pt] at (6.8,2.5) {$\alpha$};
    \draw[black] (6.8,2.1) arc[start angle=-18,end angle=21,radius=1.0];

    \draw[blue] (0.9,0) arc[start angle=0,end angle=19,radius=0.9];
    \node[blue,right] at (0.9,0.3) {$\theta$};
    
    \draw[red] (1.5,0) arc[start angle=0,end angle=5.5,radius=1.5];
    \node[red,right] at (1.5,0.1) {$\beta$};

    \draw[<->] (0,-0.7) -- node[below] {$D_L$} (6,-0.7);
    \draw[<->] (6,-0.7) -- node[below] {$D_{LS}$} (11.5,-0.7);
    \draw[<->] (0,-1.2) -- node[below] {$D_S$} (11.5,-1.2);

    \fill (O) circle (1.5pt) node[below left] {$O$};
    \fill (S) circle (1.5pt) node[right] {$S$};


\end{tikzpicture}
\caption{\textbf{Gravitational lensing geometry in the equatorial ($x$-$y$) plane:} $O$, $L$, and $S$ represent the observer, lens, and source, separated by angular-diameter distances $D_L$, $D_S$, and $D_{LS}$. Dashed lines $SY$ and $OZ$ are tangents to the null geodesic at $S$ and $O$. The perpendicular $LX$ from $L$ to $SY$ defines the impact parameter $b$, and $\alpha$ is the deflection angle.}
\label{fig:lens_geometry}
\end{figure}
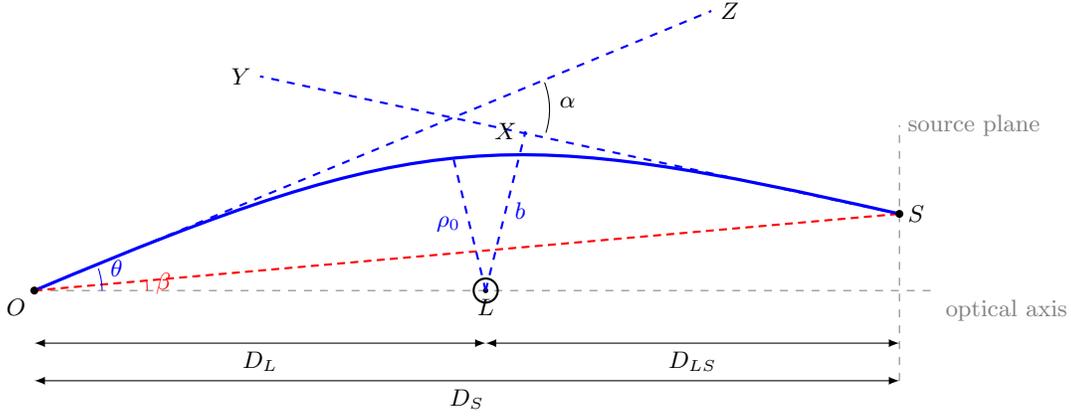

\textbf{Motivation and Background} 
The phenomenon in which a massive object deflects a light ray is called gravitational lensing (GL).
\cite{Wambsganss:1998gg,ORBi-f3f7ce61-7628-4d1a-b937-387c26218fd7, Prat:2025ucy,Bartelmann:2016dvf,Bozza_2002,Bozza:2001xd,Bozza:2002zj,Bozza:2003cp,Bozza:2008ev,Wang:2024iwt,Liu:2026yyj}. Figure~\ref{fig:lens_geometry} depicts the lensing geometry. The spacetime we consider is asymptotically flat, where both observer $O$ and source $S$ reside in flat regions, separated by a lens $L$. The line $OL$ connecting the observer and the lens defines the optical axis. A light ray emitted from a source $S$ at an angular position $\beta$ is deflected by the lens $L$ and reaches the observer $O$ at an apparent angular position $\theta$. The amount of the deflection is quantified by the deflection angle $\hat{\alpha}$.
We model the lens as a static, spherically symmetric spacetime described by the line element:
\begin{equation}
ds^2 = -A(r)dt^{2} + B(r)dr^{2} + C(r)\left(d\theta^{2}+\sin^{2}\theta d\phi^{2}\right).
\end{equation}
Spherical symmetry allows us to restrict the photon trajectories to the equatorial (paper) plane without loss of generality. 
There are two conserved quantities along the null geodesics: Energy $\mathcal{E}$ and angular momentum $\mathcal{J}$. The impact parameter $b$ is defined as
\begin{equation}
b \equiv \frac{\mathcal{J}}{\mathcal{E}} = \sqrt{\frac{C(r_0)}{A(r_0)}}\,,
\label{eq:impact_parameter}
\end{equation}
where $r_0$ is the distance of closest approach, i.e., $\dot{r}|_{r_0}=0$.
From the null geodesic equation, the deflection angle $\alpha$ is expressed as a function of $r_0$ \cite{Virbhadra:1998dy,Weinberg:1972kfs}:
\begin{equation}
\alpha(r_0) = 2\int^{\infty}_{r_0} \sqrt{\frac{B(r)}{C(r)}} \left[ \frac{C(r)A(r_0)}{C(r_0)A(r)} - 1 \right]^{-1/2} dr - \pi\;,
\label{eq:deflection_angle_generic}
\end{equation}
where 2 is the consequence of the incoming and outgoing paths being symmetric.

In the strong deflection regime, the geometric relations between the source, lens, and observer are given by the exact lens equation (often referred to as the Ohanian or Virbhadra-Ellis lens equation)~\cite{Bozza:2008ev}:
\begin{equation}
D_S \tan\beta = \frac{D_L \sin \theta - D_{LS}\sin(\alpha-\theta)}{\cos(\alpha-\theta)},
\label{eq:lens_equation}
\end{equation}
where the observer-lens, lens-source, and observe-source angular-diameter distances are denoted by $D_L$, $D_{LS}$, and $D_S = D_L + D_{LS}$, respectively.

Since the surface brightness is preserved in gravitational lensing according to Liouville’s theorem, the 
magnification $\mu$ is defined as the ratio of the solid angles of the image and the source:
\begin{equation}
\mu \equiv \frac{d\Omega_{\theta}}{d\Omega_{\beta}} = \left(\frac{\sin\beta}{\sin\theta}\frac{d\beta}{d\theta}\right)^{-1}.
\label{eq:magnification_general}
\end{equation}  
 As we will see, the lensing observables are determined by deflection $\alpha$.
 
Recent work \cite{Zhang:2025ccx} shows that broad classes of static, spherically symmetric geometries can arise as exact vacuum solutions of generally covariant gravity theories. A regular black hole can potentially be found related to vacuum solutions arising from certain generally covariant gravity theories. Gravitational lensing of a regular black hole has been studied \cite{Eiroa:2010wm,Zhao:2017cwk,Liu:2026yyj,10.1088/1674-1056/ae29fa}. 
A recent work~\cite{Calza:2025mrt} constructs Bardeen-like regular black holes that resolve the Cauchy horizon issue in the original Bardeen black hole \cite{Poisson:1990eh}. A Bardeen-like black hole \cite{Calza:2025mrt} has the metric
\begin{equation}
ds^2=-\left(1-\frac{2M(\rho)}{\rho}\right)dt^{2}
+\left(1-\frac{2M(\rho)}{\rho}\right)^{-1}d\rho^{2}
+r^2(\rho)\,d\Omega^{2}\,,
\label{eq:Bardeen_like_metric}
\end{equation}
where the mass function $M(\rho)$ and the areal radius $r(\rho)$ are defined as
\begin{equation}
M(\rho)=\frac{m\rho^{3}}{(\rho^{2}+\ell^{2})^{3/2}},
\qquad
r(\rho)=\frac{(\rho^{2}+\ell^{2})^{3/2}}{\rho^2}\, ,
\end{equation}
and satisfies the condition that $\rho \geq \rho_0$, with $\frac{\mathrm{d}r(\rho)} {\mathrm{d}\rho}|_{\rho_0}=0$. 
Here, $m$ represents the ADM mass, and $\ell$ is the regular-core (length) scale and has an upper bound $\ell < \frac{4m}{3\sqrt{3}}$. Setting $\ell=0$ recovers the standard Schwarzschild black hole. More details of Bardeen-like black holes can be found in Appendix \ref{sec:Bardeen_like_BH}.

\section{Weak-Field Lensing}
\label{sec:weak_GBT}

To investigate the lens's gravitational signature at astrophysical distances, we turn to the weak-field regime. Here, the impact parameter $b$ is significantly larger than the event horizon, allowing us to treat the deflection of light as a perturbative effect. 

In the weak field, the closest approach $r_0$ is very large, which means we can expand the deflection angle in $\frac{1}{r_0}$. Meanwhile, we change the integration variable $r$ to $\frac{r_0}{z}$, the deflection angle~\eqref{eq:deflection_angle_generic} becomes:
\begin{equation}
\alpha\left(\frac{1}{r_0}\right) = 2\int^{1}_{0} \sqrt{\frac{B\left(\frac{r_0}{z}\right)}{C\left(\frac{r_0}{z}\right)}} \left[ \frac{C\left(\frac{r_0}{z}\right)A(r_0)}{C(r_0)A\left(\frac{r_0}{z}\right)} - 1 \right]^{-1/2} \frac{r_0}{z^2}\,dz - \pi\;.
\end{equation}
After integration, we obtain:
\begin{equation}
\alpha\left(\frac{1}{r_0}\right)=\frac{4m}{r_0}-\frac{4m^2}{r_0^2}+\frac{15m^2\pi}{4r_0^2}+\frac{3l^2\pi}{4r_0^2}\;.
\label{eq:deflection_angle_1/r0}
\end{equation}
The relation between $\frac{1}{r_0}$ and $\frac{2m}{b}$ can be obtained from the Taylor expansion in $\frac{1}{r_0}$ of ~\eqref{eq:impact_parameter}:
 \begin{equation}
     \frac{2m}{b}=\frac{2m}{r_0}-\frac{2m^2}{r_0^2}-\frac{m^3+3ml^2}{r_0^3}-\frac{m^4-6m^2l^2}{r_0^4}+\frac{-5m^5+18m^3l^2+15ml^4}{4r_0^5}+\mathcal{O}\left(\frac{1}{r_0^5}\right).
 \end{equation}
We solve this transcendental equation and obtain:
\begin{equation}
    \frac{1}{r_0}=\frac{1}{b}+\frac{m}{b^2}+\mathcal{O}\left(\frac{1}{b^2}\right).
    \label{eq:r_0_b}
\end{equation}
Plugging \eqref{eq:r_0_b} into~\eqref{eq:deflection_angle_1/r0}, we obtain the deflection angle to first order correction of $\ell$:
\begin{equation}
\alpha(b) \simeq \frac{4m}{b} + \frac{15\pi}{4}\frac{m^{2}}{b^{2}} + \frac{3\pi}{4}\frac{\ell^2}{b^2} + \mathcal{O}\!\left(m^3, m^2\ell^2\right)\;,
\label{eq:deflection_angle}
\end{equation}
where the first term is the standard Einstein deflection angle for a point mass. The most notable feature of this model, however, is the sign of the $\ell$-correction term of the deflection angle. The correction involving $ \ell^2/b^2$ enters with a \textbf{positive} sign. This positive deviation distinguishes the Bardeen-like black holes from the standard Bardeen black holes. In the standard Bardeen case, the regular core typically produces a negative correction that reduces the deflection angle. In contrast, the Bardeen-like black hole enhances the deflection in the weak-field regime. The regular-core scale $\ell$ effectively deepens the gravitational potential slightly at large distances, mimicking a small increase in the effective mass.

\begin{figure}[h!]
\centering
    \begin{subfigure}{0.48\linewidth}
        \includegraphics[width=1.0\textwidth]{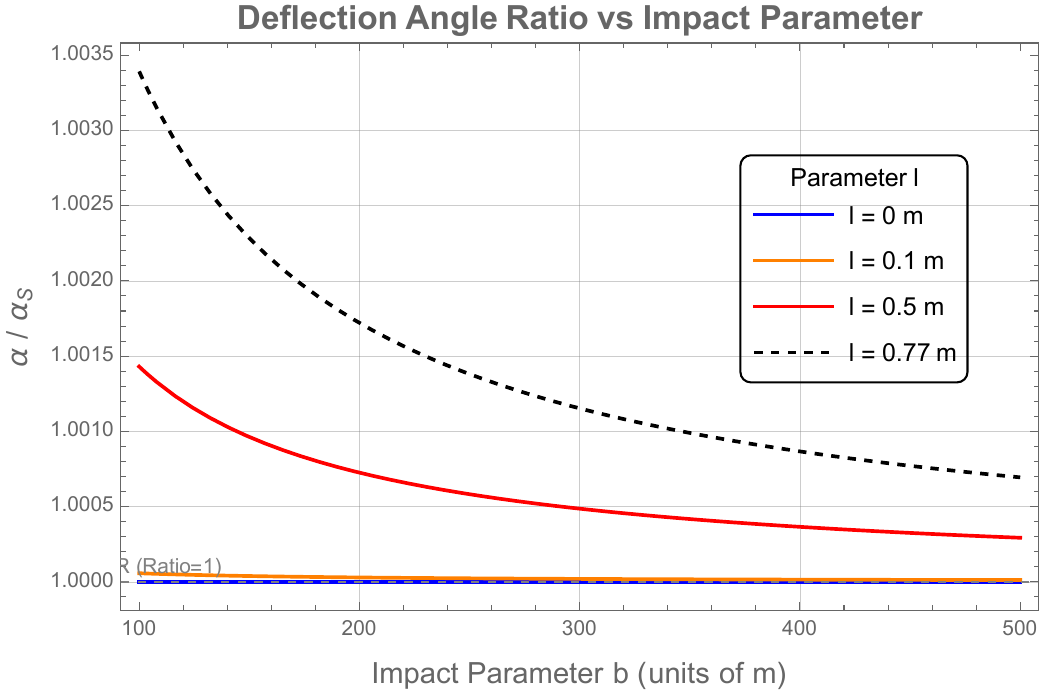}
        \caption{}
        \label{fig:deflection_ratio_bardeen}
    \end{subfigure}
    \hfill
    \begin{subfigure}{0.48\linewidth}
        \includegraphics[width=1.0\textwidth]{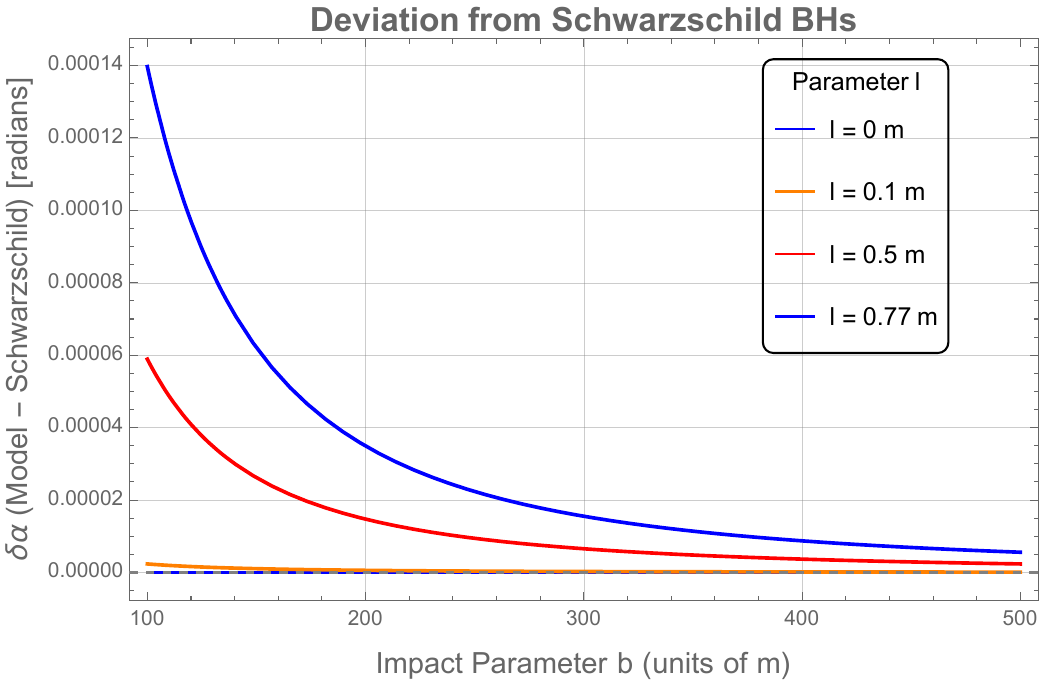}
        \caption{}
        \label{fig:deflection_Bardeen}
    \end{subfigure}
	\caption{(a) Deflection angle $\alpha$ plotted against the impact parameter $b$ for various values of the regularization parameter $\ell$. (b) Deviation from the Schwarzschild prediction ($\alpha - \alpha_{\text{Schw}}$). Positive values indicate that the Bardeen-like $\ell$ enhances deflection in the weak-field region.}
	\label{fig:deflection_weak}
\end{figure}

This behavior is shown in Fig.~\ref{fig:deflection_weak}. Figure~\ref{fig:deflection_ratio_bardeen} displays the ratio of the total deflection angle, while Fig.~\ref{fig:deflection_Bardeen} isolates the deviation from the Schwarzschild black hole. The deviation is consistently positive and grows monotonically with $\ell$. 

The image is called the Einstein ring when the source, lens, and observer are perfectly aligned ($\beta=0$). In this configuration, the lens equation simplifies to $\theta_E = \frac{D_{LS}}{D_S} \alpha(D_L \theta_E)$. Inserting our derived deflection angle, we obtain an implicit equation for the  Einstein ring:
\begin{equation}
\theta_{E} = \frac{D_{LS}}{D_{S}} \left[ \frac{4m}{D_L\theta_E} + \frac{15\pi}{4}\frac{m^{2}}{(D_L\theta_E)^{2}} + \frac{3\pi}{4}\frac{\ell^2}{(D_L\theta_E)^2} \dots \right].
\label{eq:enstein_ring_eqn}
\end{equation}
The first term corresponds to the point mass's radius, and the correction terms yield a larger ring radius. In the Appendix \ref{sec:Einstein Ring of Galaxy}, we show that the calculated Einstein ring using observational data in ESO~325-G004 is consistent with the observed value of the Einstein ring in ESO~325-G004 ~\cite{Smith:2005pq,Smith:2013ena}. The theoretical results of deflection angles and Einstein rings are consistently slightly larger than those in the Schwarzschild case. This implies that while the current model is consistent with available data, future observations with higher angular resolution may constrain $\ell$.

\section{Strong Field Limit and Lensing Observables}
\label{sec:SDL_observables}
\subsection{Lensing Observables in Strong Field Limit}
\begin{figure}[t!]
    \centering
    \includegraphics[width=0.70\textwidth]{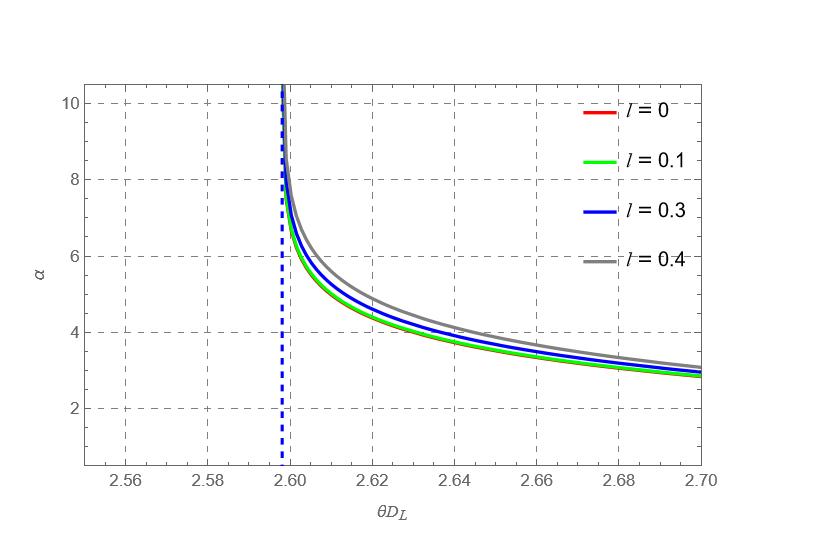}
    \caption{Variation of the deflection angle as a function of the impact parameter $u \approx \theta D_L$ (normalized to $R_s=2m$) for various values of $\ell$. Vertical dashed line marks the critical impact parameter $u_m \approx 3\sqrt{3}m$. At this limit, the deflection angle diverges logarithmically, signifying the onset of the strong lensing regime ($\alpha > 2\pi$).}
    \label{fig:SDL_deflection}
\end{figure}

To probe gravitational lensing in the vicinity of black holes, we must consider the strong field regime. In this region, the weak-field approximation breaks down as light rays are deflected near the photon sphere of a black hole. We assume that photons traversing this region follow null geodesics exterior to the photon sphere of the Bardeen-like black hole, allowing us to employ the SDL method. This method, developed by Bozza~\cite{Bozza:2001xd,Bozza:2002zj}, provides an accurate analytic framework for describing the relativistic images formed by photons that wind multiple times around the black hole before reaching the observer. 

For convenience, we rewrite the Bardeen-like black hole metric \eqref{eq:Bardeen_like_metric} as:
\begin{equation}
    ds^2 = -\left(1-\frac{2m}{r(\rho)}\right)dt^2 + \left(1-\frac{2m}{r(\rho)}\right)^{-1}d\rho^2 + r(\rho)^2 d\Omega^2,
\end{equation}
where the areal radius is defined as $r(\rho)=\frac{(l^2+\rho^2)^{3/2}}{\rho^2}$. Meanwhile, we set \begin{equation}
    A(\rho)\equiv1-\frac{2m}{r(\rho)},\;\;\;
    B(\rho)\equiv\left(1-\frac{2m}{r(\rho)}\right)^{-1},\;\;
    C(\rho)\equiv r(\rho).
    \label{eq:Bardeen-Like rewrite}
\end{equation}

Near a black hole, gravity is strong enough to force photons into unstable circular orbits. The collection of such circular orbits is called the black hole's photon sphere, whose radius is determined by the root of the equation:
\begin{equation}
    \frac{A'(\rho_{ps})}{A(\rho_{ps})} = \frac{C'(\rho_{ps})}{C(\rho_{ps})}.
\end{equation}
Solving this for the Bardeen-like black hole yields the critical condition:
\begin{equation}
    r(\rho_{ps}) = 3m,
\end{equation}
where $\rho_{ps}$ is the largest real root of the equation
$
    (\rho_{ps}^2+l^2)^{3/2} = 3m\rho_{ps}^2.
$
Consequently, we can define a critical impact parameter $u_{ps}$. Rays with impact parameters $u < u_{ps}$ are captured by the black hole, contributing to the black hole's dark ``shadow'', while rays with $u > u_{ps}$ escape. The boundary between the shadow and the photon sphere is given by:
\begin{equation}
    u_{ps} = \sqrt{\frac{C(\rho_{ps})}{A(\rho_{ps})}}= 3\sqrt{3}m.
\end{equation}
Remarkably, this result implies that the critical impact parameter depends only on the value of $r(\rho)$ at the photon sphere, which remains effectively $3m$ regardless of the regular-core scale $\ell$.

In the strong deflection limit ($u \to u_{ps}$), the deflection angle $\alpha(u)$ does not remain small; instead, it diverges logarithmically as the photon spends more time winding around the black hole~\cite{Bozza:2002zj, Bozza:2001xd,Tsukamoto:2016jzh}:
\begin{equation}
    \alpha(u) = -\bar{a} \log\left(\frac{u}{u_{ps}}-1\right) + \bar{b} + \mathcal{O}(u-u_{ps}),
    \label{eq:SDL_deflection}
\end{equation}
where $\bar{a}$ and $\bar{b}$ are called SDL coefficients. The coefficient $\bar{a}$ has a closed form:
\begin{equation}
    \bar{a} = \sqrt{\frac{2A(\rho_{ps})B(\rho_{ps})}{C''(\rho_{ps})A(\rho_{ps})-A''(\rho_{ps})C(\rho_{ps})}} = \left( \frac{\rho_{ps}}{3\sqrt{\rho_{ps}^2+l^2}-6m} \right),
\end{equation}
 but we were unable to derive an analytical expression for $\bar{b}$; it can only be evaluated numerically. In the limit $\ell \to 0$, we recover the Schwarzschild values $\bar{a}=1$ and $\bar{b} \approx -0.40023$. 

\begin{figure}[h!]
    \centering
    \begin{subfigure}[b]{0.49\linewidth}
        \centering
    \includegraphics[width=\textwidth]{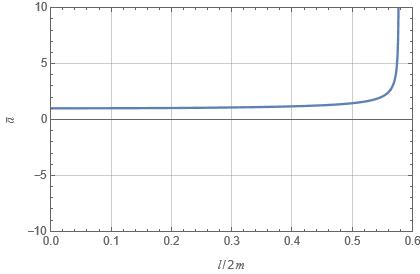}
        \caption{}
        \label{fig:abar_Bardeenlike}
    \end{subfigure}
    \hfill
    \begin{subfigure}[b]{0.49\linewidth}
        \centering
\includegraphics[width=\textwidth]{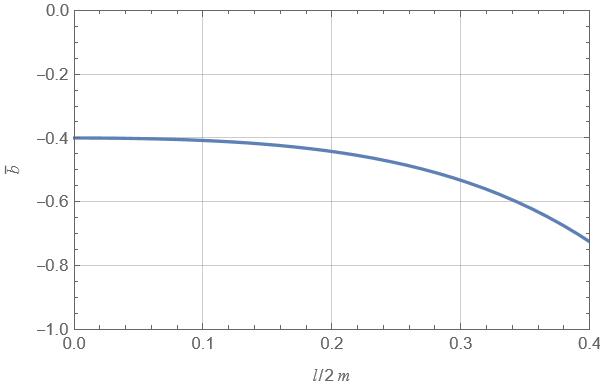}
        \caption{}
        \label{fig:bbar_Bardeenlike}
    \end{subfigure}
    \caption{Behavior of the strong lensing coefficients with respect to the regular parameter $\ell$. Panel (a) shows the increase in $\bar{a}$, while Panel (b) shows the decrease in $\bar{b}$.}
    \label{fig:lensing_coefficients}
\end{figure}

As depicted in Fig.~\ref{fig:lensing_coefficients} and detailed in Table~\ref{table:table_parameter}, $\bar{a}$ monotonically increases with $\ell$, whereas $\bar{b}$ decreases as $\ell$ increases. The regular-core scale $\ell$ of the Bardeen-like black hole modifies the lensing signature by altering the SDL coefficients.

\begin{table}[h!]
    \centering
    \caption{Numerical estimates for the strong lensing coefficients ($\bar{a}, \bar{b}$) and the critical impact parameter $u_{ps}$, normalized to the Schwarzschild radius $R_s=2m$. Case $\ell=0$ corresponds to the Schwarzschild limit.}
    \label{table:table_parameter}
    \setlength{\tabcolsep}{8pt}
    \renewcommand{\arraystretch}{1.2}
    \begin{tabular}{l c c c}
        \hline\hline
        $\ell/2m$ & $\bar{a}$ & $\bar{b}$ & $u_{ps}/R_s$ \\
        \hline
        $\ell=0$    & 1.00000 & -0.40023  & 2.59808 \\
        $\ell=0.1$  & 1.00683 & -0.407874 & 2.59808 \\
        $\ell=0.3$  & 1.07636 & -0.532713 & 2.59808 \\
        $\ell=0.4$  & 1.17461 & -0.72538  & 2.59808 \\
        \hline\hline
    \end{tabular}
    
\end{table}

The divergence of the deflection angle near $u_{ps}$ (shown in Fig.~\ref{fig:SDL_deflection}) implies that photons can loop around the black hole multiple times ($n$) before reaching the observer. Each full loop ($2\pi$) creates a new ``relativistic image''. Although the total deflection angle $\alpha$ is large ($\alpha > 2\pi$), the observed angular separation of these images from the optical axis is small as both source and observer far away from the lens. This physical setup justifies the use of the small-angle approximation ($\beta, \theta \ll 1$) in the lens equation~\cite{Bozza:2008ev,Virbhadra:1999nm}:
\begin{equation}
    \beta = \theta - \frac{D_{LS}}{D_S}\Delta\alpha_n,
    \label{eq:SDL_lens_eq}
\end{equation}
where $\Delta\alpha_n = \alpha(\theta)-2n\pi  $. Solving Eq.~\eqref{eq:SDL_lens_eq} using the expansion in Eq.~\eqref{eq:SDL_deflection}, we obtain the position of the $n$-th relativistic image by \cite{Bozza_2002}:
\begin{equation}
    \theta_{n} \simeq \theta_{n}^{0} + \frac{u_{ps}e_n(\beta-\theta_{n}^{0})D_{S}}{\bar{a}D_{LS}D_{L}},
    \label{eq:position_SDL_image}
\end{equation}
where $e_n = \exp\left(\frac{\bar{b}-2n\pi}{\bar{a}}\right)$, and $\theta_{n}^{0} = u_{ps}(1+e_n)/D_L$ represents the theoretical image position for a perfect $2n\pi$ loop. 

The magnification of these images can be obtained by applying the small-angle approximation to Eq.~\eqref{eq:magnification_general}, the magnification of the $n$-th image is found to be~\cite{Bozza:2002zj}:
\begin{equation}
    \mu_n = \left. \left( \frac{\beta}{\theta} \frac{d\beta}{d\theta} \right)^{-1} \right|_{\theta^0_n} = \frac{u^2_{ps}e_n(1+e_n)D_S}{\bar{a}\beta D_{LS}D^2_{L}}.
    \label{eq:SDL_magnification}
\end{equation}
Since $e_n$ decays exponentially with $n$, the brightness of the relativistic images drops rapidly. This implies that only the first relativistic image ($n=1$) is likely to be detectable, appearing as a faint, thin ring closely hugging the black hole's shadow.

Finally, we can find three strong lensing SDL observables~\cite{Bozza:2001xd}:
\begin{itemize}
    \item $\theta_{\infty} = u_{ps} / D_L$: The asymptotic position of relativistic images.
    \item $s = \theta_1 - \theta_{\infty}$: The angular separation between the first relativistic image and the others. This effectively measures the ``thickness'' between the first relativistic image and the others.
    \item $r_{\mathrm{mag}}$: The relative flux ratio between the first image and the aggregate flux of all subsequent images ($\sum_{n=2}^{\infty}\mu_n$). This quantifies the dominance of the outermost ring.
\end{itemize}
These are explicitly calculated as\cite{Bozza_2002}:
\begin{equation}
    s \approx \theta_{\infty} \exp{\left(\frac{\bar{b}-2\pi}{\bar{a}}\right)}, \qquad 
    r_{\mathrm{mag}} \approx \frac{5\pi}{\bar{a}\ln{10}}.
\end{equation}
Beyond purely geometric parameters, the time delay between two relativistic images 
offers a powerful diagnostic tool for probing the strong-field regime. This time delay 
originates from the additional loops photons execute around the photon sphere before 
escaping toward the observer. In the context of SDL and 
assuming a near-perfect alignment ($\beta \approx 0$), the delay between the $n$-th 
and $m$-th order images can be expressed as \cite{Bozza:2003cp, Lu:2016gsf}:
\begin{equation}
\Delta T_{n,m} \simeq 2\pi(n-m)u_{ps} + 2\sqrt{\frac{B_{ps} u_{ps}}{A_{ps} \hat{c}}} 
\exp\left(\frac{\bar{b}}{2\bar{a}}\right) \left[ \exp\left(-\frac{m\pi}{\bar{a}}\right) 
- \exp\left(-\frac{n\pi}{\bar{a}}\right) \right].
\label{eq:time_delay}
\end{equation}
Here, the primary contribution to the delay is governed by the impact parameter at the 
photon sphere, $u_{ps}$, while the finer corrections are determined by the 
SDL coefficients $\bar{a}$ and $\bar{b}$, as well as the metric-dependent 
factor $\hat{c}$ \cite{Bozza_2002} expressed as
\begin{equation}
\hat{c} = \hat{\beta}_{ps} \sqrt{\frac{A_{ps}}{C_{ps}^3}} \frac{(C'_{ps})^2}{2(1-A_{ps})^2}.
\end{equation}

In the following subsection, we apply this formalism to the supermassive black hole candidates M87* and Sgr A*.

 \subsection{Lensing observables of the supermassive black holes Sgr~A* and M87*} 

 Assuming the supermassive black holes Sgr~A* and M87* being Bardeen-like, we estimate and compare their lensing observables with those of the Schwarzschild black hole.
The estimated mass of M87* is $(6.5\pm 0.7)\times 10^9 M_\odot$ at a distance of $16.8$ Mpc~\cite{EventHorizonTelescope:2019pgp,EventHorizonTelescope:2019ggy}.
For Sgr~A*, the estimated mass is $4.28^{\pm0.21}_{\pm0.1}\times 10^6 M_{\odot}$ at a distance of $8.32^{\pm 0.07}_{\pm 0.14}$ kpc~\cite{EventHorizonTelescope:2022apq,EventHorizonTelescope:2022wkp}.

We calculated the angular positions of the first and second relativistic images for Sgr~A* and M87* using Eq.~\eqref{eq:position_SDL_image}.
Our numerical results shows that relative flux ratio $r_{\mathrm{mag}}$ is a function of $\ell$ that deviates from the Schwarzschild case.

\begin{figure}[h!]
\centering
    \begin{subfigure}{0.49\linewidth}
        \includegraphics[width=1.0\textwidth]{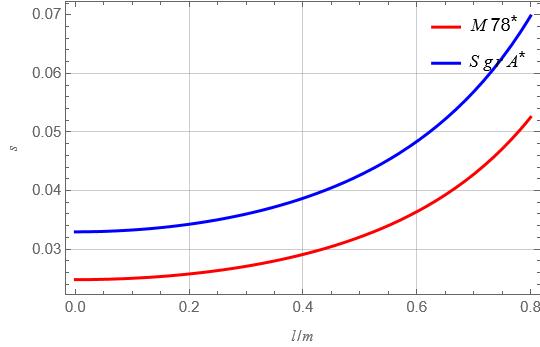}
        \caption{}
        \label{fig:s_SDL}
    \end{subfigure}
    \begin{subfigure}{0.49\linewidth}
        \includegraphics[width=1.0\textwidth]{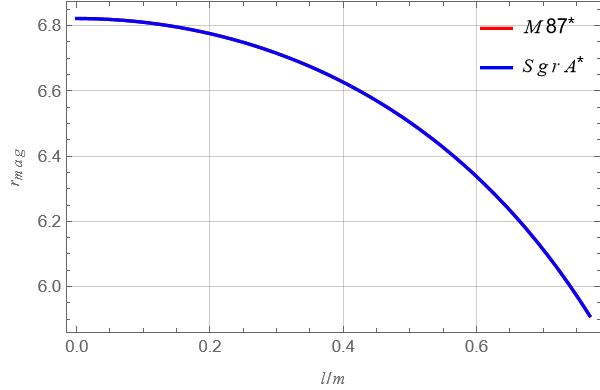}
        \caption{}
        \label{fig:rm_SDL}
    \end{subfigure}
	\caption{Variation of lensing observables with $\ell$ for Sgr~A* (blue) and M87* (red). (a) The angular separation $s$ increases with $\ell$. (b) The magnitude difference $r_{\mathrm{mag}}$ decreases with $\ell$. Note that the mass $m$ is absorbed in the parameter $\ell$ and $\rho$, which both are in units of mass, so the curves for both sources overlap.}
	\label{fig:lensing_observables}
\end{figure}

\begin{figure}[h!]
\centering
    \begin{subfigure}{0.49\linewidth}
        \includegraphics[width=1.0\textwidth]{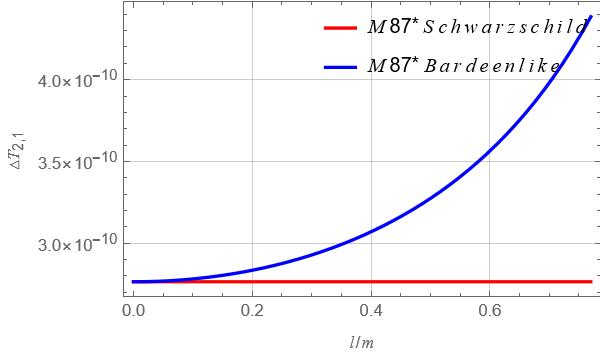}
        \caption{}
        \label{fig:timedelay_SDL_M87_second}
    \end{subfigure}
    \begin{subfigure}{0.49\linewidth}
        \includegraphics[width=1.0\textwidth]{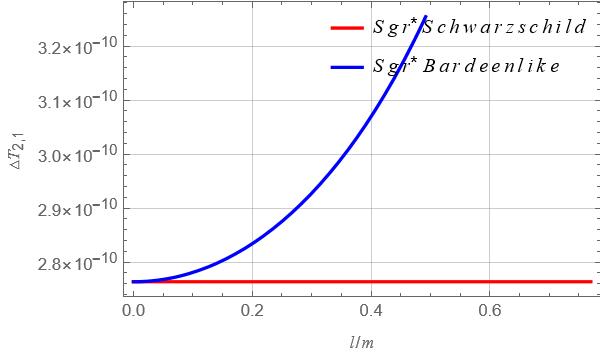}
        \caption{}
        \label{fig:timedelay_SDL_sgr_second}
    \end{subfigure}
	\caption{Deviation of time delay from Schwarzschild black hole between the first and second relativistic images and varies with $\ell$  for M87* (a) and Sgr~A* (b), both are in units of seconds. Both deviation increases with $\ell$.}
	\label{fig:lensing_timedelay}
\end{figure}

According to our numerical results in Table~\ref{table:table_SDL_observables} and Table~\ref{table:table_deviation_observables}, the angular separation $s$ increases with $\ell$ and ranges from about 33 to 65 nanoarcseconds (nas) for Sgr~A* and from 25 to 49 nanoarcseconds (nas) for M87*.
As shown in Fig.~\ref{fig:s_SDL}, as $\ell$ increases, the separation between the first relativistic image and the others grows.
This behavior follows the same pattern as observed in standard Bardeen black holes~\cite{Eiroa:2010wm}.
In contrast, the relative flux ratio $r_{\mathrm{mag}}$ is smaller than that in the Schwarzschild limit and decreases as $\ell$ increases, taking the same values for both Sgr~A* and M87* as seen in Fig.~\ref{fig:rm_SDL}, ranging from 6.82 to 5.91. This behavior also follows the same pattern as observed in standard Bardeen black holes~\cite{Eiroa:2010wm}.
The $\theta_{\infty}$ remains constant at $19.83\,\mu\text{as}$ for Sgr~A* and $26.55\,\mu\text{as}$ for M87*,   showing no deviation from the Schwarzschild prediction, but differs from the standard Bardeen prediction, where $\theta_{\infty}$ typically decreases as $\ell$ increases \cite{Eiroa:2010wm}. While we primarily focus on the relative flux ratio $r_{\mathrm{mag}}$ to characterize the lensing signatures, the magnifications ($\mu_n$) and angular positions ($\theta_n$) of the individual images depend much on the source position $\beta$. Tables \ref{table:table_magnification_Sgr_Ba} and \ref{table:table_magnification_M87_Ba} show the magnifications and image positions of the first two relativistic images for Sgr~A* and M87*, respectively. Figure \ref{fig:lensing_timedelay} depicts the total time delay $\Delta T_{2,1}$ between the first and second relativistic images using the observational data for M87* and Sgr~A*. For both supermassive black holes, the total time delay increases with $\ell$. For M87*, this increment is so small that it cannot be visually discerned on the primary scale. This increasing behavior and the deviation from the standard Schwarzschild prediction can be better observed in Fig.~\ref{fig:timedelay_SDL_M87_second} and Fig.~\ref{fig:timedelay_SDL_sgr_second}, respectively, where these deviations are from the second term of Eq.~\eqref{eq:time_delay}. 
\begin{table}[ht]
    \centering
    \setlength{\tabcolsep}{5pt}
    \renewcommand{\arraystretch}{1.3}
    
    \caption{Lensing observables for M$87^*$ and Sgr $A^*$. The row $\ell=0$ represents the Schwarzschild reference values.}
    \label{table:table_SDL_observables}
    
    \begin{tabular}{l c c c c c}
        \hline\hline
         & \multicolumn{2}{c}{\textbf{M87*}} & \multicolumn{2}{c}{\textbf{Sgr A*}} & \\
        $\ell/m$ & $\theta_{\infty} (\mu\text{as})$ & $s (\mu\text{as})$ & $\theta_{\infty} (\mu\text{as})$ & $s (\mu\text{as})$ & $r_{\mathrm{mag}}$ \\
        \hline
        0.0  & 19.8012 & 0.02478 & 26.3274 & 0.03295 & 6.822 \\
        0.1  & 19.8012 & 0.02502 & 26.3274 & 0.03327 & 6.810 \\
        0.3  & 19.8012 & 0.02707 & 26.3274 & 0.03599 & 6.716 \\
        0.5  & 19.8012 & 0.03206 & 26.3274 & 0.04263 & 6.504 \\
        0.77 & 19.8012 & 0.04912 & 26.3274 & 0.06532 & 5.911 \\
        \hline\hline
    \end{tabular}
    
    \vspace{0.5cm}
    
    \caption{Deviations from the Schwarzschild limit ($\ell=0$) for the lensing observables.}
    \label{table:table_deviation_observables}
    
    \begin{tabular}{l c c c c c}
        \hline\hline
         & \multicolumn{2}{c}{\textbf{M87*}} & \multicolumn{2}{c}{\textbf{Sgr A*}} & \\
        $\ell/m$ & $\delta\theta_{\infty} (\mu\text{as})$ & $\delta s (\mu\text{as})$ & $\delta\theta_{\infty} (\mu\text{as})$ & $\delta s (\mu\text{as})$ & $\delta r_{\mathrm{mag}}$ \\
        \hline
        0.0  & 0 & 0           & 0 & 0           & 0 \\
        0.1  & 0 & $+0.00024$  & 0 & $+0.00032$  & $-0.011$ \\
        0.3  & 0 & $+0.00229$  & 0 & $+0.00304$  & $-0.106$ \\
        0.5  & 0 & $+0.00728$  & 0 & $+0.00968$  & $-0.318$ \\
        0.77 & 0 & $+0.02434$  & 0 & $+0.03237$  & $-0.911$ \\
        \hline\hline
    \end{tabular}
\end{table}
\begin{table}[ht]
    \centering
    \setlength{\tabcolsep}{5pt}
    \renewcommand{\arraystretch}{1.3}
    
\caption{Magnifications and image positions of the first two relativistic images due to lensing by Sgr A* for $\beta=1$ with $D_{LS}/D_S=1/2$.}
    \label{table:table_magnification_Sgr_Ba}
    
    \begin{tabular}{l c c c c c c c c}
        \hline\hline
         & & & & \multicolumn{2}{c}{\textbf{Sgr A*}} & \\
         $\ell/m$ & $\mu_{p,1}(10^{-11})$ & $\mu_{p,2}(10^{-14})$ & $\mu_{s,1}(10^{-11})$  & $\mu_{s,2}(10^{-14})$ & $\theta_{p,1} (\mu\text{as})$ & $\theta_{p,2} (\mu\text{as})$ & $\theta_{s,1} (\mu\text{as})$ & $\theta_{s,2} (\mu\text{as})$ \\
        \hline
           0.0 & $3.43$ & $6.38$ & $-3.43$ & $-6.38$ & 26.4610 & 26.3276 & 26.4610 & 26.3276 \\
         0.1 & $3.46$ & $6.50$ & $-3.46$ & $-6.50$ & 26.4624 & 26.3276 & 26.4624 & 26.3276\\
         0.3 & $3.73$ & $7.64$ & $-3.73$ & $-7.64$ & 26.4747 & 26.3277 & 26.4747 & 26.3277 \\
         0.5 & $4.39$ & $10.9$ & $-4.39$ & $-10.9$ & 26.5065 & 26.3278 & 26.5065 & 26.3278\\
         0.77 & $6.93$ & $29.6$ & $-6.93$ & $-29.6$ & 26.6367 & 26.3287 & 26.6367 & 26.3287\\
        \hline\hline
    \end{tabular}

    \vspace{0.5cm}
    
    \caption{Magnifications and image positions of the first two relativistic images due to lensing by M87* for $\beta=1$ with $D_{LS}/D_S=1/2$.}
    \label{table:table_magnification_M87_Ba}
    
    \begin{tabular}{l c c c c c c c c}
        \hline\hline
         & & & & \multicolumn{2}{c}{\textbf{M87*}} & \\
         $\ell/m$ & $\mu_{p,1}(10^{-12})$ & $\mu_{p,2}(10^{-15})$ & $\mu_{s,1}(10^{-12})$  & $\mu_{s,2}(10^{-15})$ & $\theta_{p,1} (\mu\text{as})$ & $\theta_{p,2} (\mu\text{as})$ & $\theta_{s,1} (\mu\text{as})$ & $\theta_{s,2} (\mu\text{as})$ \\
        \hline
           0.0 & $4.77$ & $8.90$ & $-4.77$ & $-8.90$ & 19.8259 & 19.8012 & 19.8259 & 19.8012 \\
         0.1 & $4.81$ & $9.06$ & $-4.81$ & $-9.06$ & 19.8262 & 19.8012 & 19.8262 & 19.8012\\
         0.3 & $5.13$ & $10.6$ & $-5.13$ & $-10.6$ & 19.8282 & 19.8012 & 19.8282 & 19.8012 \\
         0.5 & $5.89$ & $14.7$ & $-5.89$ & $-14.7$ & 19.8332 & 19.8012 & 19.8332 & 19.8012\\
         0.77 & $8.20$ & $35.3$ & $-8.20$ & $-35.3$ & 19.8503 & 19.8014 & 19.8503 & 19.8014\\
        \hline\hline
    \end{tabular}
    
\end{table}
The angular diameter of the shadow of M87* is $\theta_{\mathrm{sh}}=42\pm3\,\mu\text{as}$~\cite{EventHorizonTelescope:2019dse,EventHorizonTelescope:2019pgp,EventHorizonTelescope:2019ggy}, based on a mass $M=(6.50\pm 0.7)\times 10^9 M_{\odot}$ and distance $D_L=16.8$ Mpc. We use the apparent radius of the photon sphere $\theta_{\infty}$ as a proxy for the angular size of the black hole shadow. As shown in Table~\ref{table:table_SDL_observables}, $\theta_{\infty}=19.8012\,\mu\text{as}$, which is $\ell$-independent. The resultant shadow diameter is $\theta_{\mathrm{sh}}=2\theta_{\infty}=39.6024\,\mu\text{as}$, which falls well within the EHT's observed interval and the error bars.

Similarly, in 2022, the EHT observed Sgr~A* at the center of our Milky Way, revealing a shadow image with a diameter of $\theta_{\mathrm{sh}}=48.7\pm 7\,\mu\text{as}$~\cite{EventHorizonTelescope:2022wkp}. This observation suggests a black hole mass of $M=4.0^{+1.1}_{-0.6}\times 10^6 M_{\odot}$ and a Schwarzschild shadow deviation $\delta=-0.08^{+0.09}_{-0.09}$ (VLTI), $-0.04^{+0.09}_{-0.10}$ (Keck). These findings provide a new method to probe the strong-field region of Sgr~ A*~\cite{EventHorizonTelescope:2022apq,EventHorizonTelescope:2022wkp, EventHorizonTelescope:2022urf}. The EHT collaboration used three independent algorithms to determine that the mean measured value of the shadow angular diameter $\theta_{\mathrm{sh}}$ falls within the range $(46.9,50.0)\,\mu\text{as}$ and within $(41.7,55.7)\,\mu\text{as}$ for $1\sigma$ interval~\cite{EventHorizonTelescope:2022xqj,EventHorizonTelescope:2022wkp}. From Table~\ref{table:table_SDL_observables}, the value for Sgr~A* is $\theta_{\infty}=26.3274\,\mu\text{as}$, implying a shadow diameter $2\theta_{\infty}=52.6548\,\mu\text{as}$ for the Bardeen-like black hole, which indeed falls well within the EHT's observed interval and the error bars.
In practice, determining the regular-core scale $\ell$ of the Bardeen-like black hole requires either separation $s$ or the relative flux ratio $r_{\mathrm{mag}}$.

\begin{figure}[h!]
\centering
    \begin{subfigure}{0.49\linewidth}
        \includegraphics[width=1.0\textwidth]{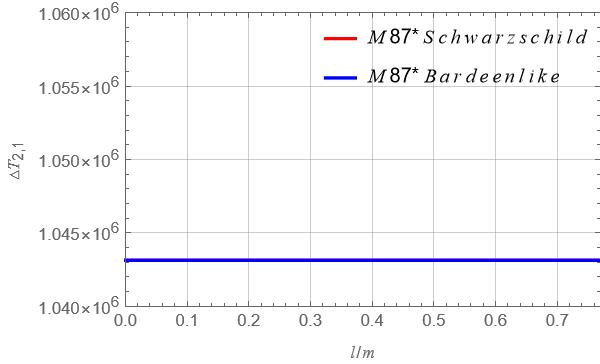}
        \caption{}
        \label{fig:timedelay_SDL_M87}
    \end{subfigure}
    \begin{subfigure}{0.49\linewidth}
        \includegraphics[width=1.0\textwidth]{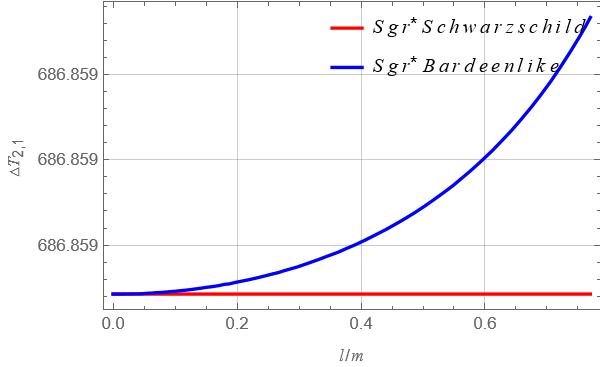}
        \caption{}
        \label{fig:timedelay_SDL_sgr}
    \end{subfigure}
	\caption{Total Time delay between the first and second relativistic images and varies with $\ell$  for M87* (a) and Sgr~A* (b), both are in units of seconds. (a) The time delay increases with $\ell$, however, the changes coming from the second term is negligible compared with the first term in Eqn. \eqref{eq:time_delay}. (b) The time delay increases with $\ell$.}
	\label{fig:lensing_timedelay}
\end{figure}

\section{Conclusion and Discussion}

In this work, we first investigated gravitational lensing in the weak-field regime. We found that the deflection angle increases with $\ell$, exhibiting a positive deviation. This contrasts the standard Bardeen black hole, which produces a negative deviation. To assess the observational viability of the Bardeen-like black hole, we compared our theoretical predictions with data from the Einstein ring ($\theta_{E}$) of the elliptical galaxy ESO~325-G004. In this calculation, the total lens mass was assumed to comprise a central Bardeen-like black hole, as well as the dark and visible matter of the galaxy. The predicted $\theta_{E}$ for the Bardeen-like black hole falls within the $1\sigma$ observational uncertainty of $2.85\pm 0.40$ arcsec.

Moving to the strong-field regime, we adopted the SDL approach. The SDL coefficient $\bar{a}$ was derived analytically, while $\bar{b}$ was computed numerically. This allowed us to calculate the strong-field deflection angle and key lensing observables, including the limiting angular position of the relativistic image sequence $\theta_{\infty}$, the angular separation $s$ between relativistic images, and the relative flux ratio $r_{\mathrm{mag}}$. We showed that the strong-field deflection angle is larger than that in the Schwarzschild case and increases monotonically with $\ell$ for a fixed impact parameter.

To explore the preliminary detectability of these strong-field features, we modeled the supermassive black hole candidates Sgr~A$^\ast$ and M87$^\ast$ as Bardeen-like black holes and estimated their expected SDL observables. Notably, the Bardeen-like black hole possesses the exact same asymptotic angular position $\theta_{\infty}$ as the Schwarzschild black hole. Based solely on $\theta_{\infty}$, however, future astronomical facilities achieving an angular resolution of roughly $100\,\mathrm{nas}$ (nanoarcseconds) or better could distinguish the Bardeen-like black hole from the standard Bardeen black hole.

Furthermore, if the observational resolution can reach $\sim 10\,\mathrm{nas}$, the first relativistic image could in principle be resolved from the tightly packed inner photon ring. This would enable the direct measurement of the observables $s$ and $r_{\mathrm{mag}}$. Measuring these parameters is crucial for helping constrain the Bardeen-like model and potentially differentiating it from the classical Schwarzschild solution. Because regular black holes can mathematically relate to vacuum solutions of generally covariant gravity theories~\cite{Zhang:2025ccx}, such high-resolution observations offer an exciting avenue to test fundamental alternative theories of gravity in the strong-field regime. Achieving such an extremely high resolution, however, is far beyond the capabilities of current technologies. With future high-precision observations, especially from the future ngEHT, these constraints could be further refined, potentially differentiating Bardeen-like black holes from Schwarzschild solutions. In fact, recent work~\cite{Boos:2025nzc} shows that the microlensing of regular black holes yields a higher magnification, which may present a more viable target for near-future astronomical observations. 

In addition to the standard SDL observables $\theta_{\infty}$, $s$, and $r_{\mathrm{mag}}$, we also computed the signed magnifications and angular positions of the first two relativistic images, $(\mu_{p,1},\mu_{p,2},\mu_{s,1},\mu_{s,2})$ and $(\theta_{p,1},\theta_{p,2},\theta_{s,1},\theta_{s,2})$, together with the differential time delay $\Delta T_{2,1}$ between the first two relativistic images. 
 In this sense, the observables naturally separate into a set, $\{\theta_{\infty}, s, r_{\mathrm{mag}}, \theta_{p,i}, \theta_{s,i}\}$, mainly limited by angular resolution and dynamic range, and a observable, $\Delta T_{2,1}$, which may admit a different interpretation if the source is a compact time-variable radio source. In the geometric-optics regime, the same strong-field trajectories that generate relativistic images also apply to radio emission. Therefore, if the source is a compact radio emitter, $\Delta T_{2,1}$ could in principle be interpreted as the lag between two delayed signal copies, with relative amplitudes related by the corresponding image magnifications.

From this perspective, the quantities computed here may provide a useful phenomenological basis for future radio-timing studies in strong-gravity environments, including possible SKA-related applications. In the present Bardeen-like case, however, the deviation of $\Delta T_{2,1}$ from the Schwarzschild value is so small that its role should be regarded primarily as conceptual rather than as a practically accessible discriminator at the current stage. Nevertheless, astronomical observations of gravitational lensing in both the weak and strong-field regimes remain a potential future opportunity to search for and detect regular black holes.

\begin{acknowledgements}
The authors thank Liang Dai for his introducing and inviting us to the area of gravitational lensing and for his insightful comments on our work. The authors also thank Huanyuan Shan, Long Wang, Le Zhang, Yi Zheng, and Zhiqi Huang for their helpful comments. YW is supported by NSFC Grant No. 12475001, the Shanghai Municipal Science and Technology Major Project (Grant No. 2019SHZDZX01), Science and Technology Commission of Shanghai Municipality (Grant No. 24LZ1400100), and the Innovation Program for Quantum Science and Technology (No. 2024ZD0300101). YW is grateful for the hospitality of the Perimeter Institute during his visit, where this work was partially done. This research was supported in part by the Perimeter Institute for Theoretical Physics. Research at Perimeter Institute is supported by the Government of Canada through the Department of Innovation, Science and Economic Development and by the Province of Ontario through the Ministry of Research, Innovation and Science. 
\end{acknowledgements}

\appendix
\renewcommand\thesection{\Alph{section}}


\section{Bardeen-like Black Holes}
\label{sec:Bardeen_like_BH}
Recent work~\cite{Calza:2025mrt} provides a method to construct non-singular black holes with spherical horizon topology and no Cauchy horizon. The metric can be written as
\begin{equation}
ds^2=-\left(1-\frac{2M(\rho)}{\rho}\right)dt^{2}
+\left(1-\frac{2M(\rho)}{\rho}\right)^{-1}d\rho^{2}
+r^2(\rho)\,d\Omega^{2}\,,
\label{eq:Hayward_like_metric}
\end{equation}
where
\begin{equation}
M(\rho)=\frac{m\,\rho^{3}}{(\rho^{2}+\ell^{2})^{3/2}},
\qquad
r(\rho)
=\frac{(\rho^2+l^2)^{3/2}}{\rho^2}\,.
\end{equation}
Here $\ell$ is a length scale controlling the regularization, and $m$ is the mass parameter (in geometric units). The spacetime described by~
\eqref{eq:Hayward_like_metric} is referred to as a Hayward-like regular black hole. Setting $\ell=0$ reduces~\eqref{eq:Hayward_like_metric} to the Schwarzschild metric.

As $\rho\to\infty$, one has
\begin{equation}
1-\frac{2M(\rho)}{\rho}\to 1-\frac{2m}{\rho},
\qquad
r(\rho)\to \rho,
\end{equation}
so the geometry approaches an asymptotically Schwarzschild region. In the opposite limit $\rho\to 0^{+}$,
\begin{equation}
1-\frac{2M(\rho)}{\rho}\to 1-
\frac{\rho^{2}}{\ell^{2}},
\qquad
r(\rho)\sim \frac{2m\ell^{2}}{\rho^{2}}\to \infty,
\end{equation}
which shows that $\rho\to 0^{+}$ again corresponds to large areal radius. Using $r(\rho)\sim 2m\ell^{2}/\rho^{2}$ for small $\rho$, we further obtain
\begin{equation}
1-\frac{\rho^{2}}{\ell^{2}}
\sim 1-\frac{2m}{r}\,,
\end{equation}
so this second asymptotic region is Schwarzschild-like as well (in terms of the areal radius $r$).

The function $r(\rho)$ attains its minimum at
\begin{equation}
\frac{dr}{d\rho}=1-\frac{4m\ell^{2}}{\rho^{3}}=0
\quad\Longrightarrow\quad
\rho_{0}=(4m\ell^{2})^{1/3},
\qquad
r_{0}=r(\rho_{0})=\frac{3}{2^{1/3}}(m\ell^{2})^{1/3}.
\end{equation}
Following~\cite{Calza:2025mrt}, one chooses the branch $\rho^{3}\ge 4m\ell^{2}$ so that $r(\rho)$ is monotonic on the physical domain. The spacetime therefore connects two asymptotically Schwarzschild regions and contains a minimal two-sphere at $\rho=\rho_{0}$.

Two surfaces are of particular interest. The first is determined by the (outer) horizon condition,
\begin{equation}
1-\frac{2M(\rho_H)}{\rho_H}=0
\quad\Longleftrightarrow\quad
\rho_H^{3}+2m\ell^{2}=2m\rho_H^{2}.
\label{eq:horizon_rho}
\end{equation}
On the physical branch, one finds $r(\rho_H)=2m$. Requiring the horizon to lie outside the minimal sphere, $r_H>r_0$, yields the bound
\begin{equation}
\ell<\frac{4m}{3\sqrt{3}}\qquad\text{equivalently}\qquad4m>3\sqrt{3}\,\ell,
\end{equation}
in agreement with~\cite{Calza:2025mrt}. The second surface is the branch point
\begin{equation}
\rho_\ell^{3}=4m\ell^{2},
\end{equation}
which acts as an inner boundary of the coordinate domain. Since we restrict to $\rho^{3}\ge 4m\ell^{2}$, this surface effectively replaces an inner (Cauchy) horizon as the inner boundary of the spacetime described 
by~\eqref{eq:Hayward_like_metric}. Remarkably, the presence of a minimal areal radius and the branch restriction removes the central singularity and avoids an inner Cauchy horizon within the physical domain at the branch surface.

\section{The Einstein Ring of Galaxy ESO 325-G004}
\label{sec:Einstein Ring of Galaxy}
We use observational data from the Einstein ring of galaxy ESO~325-G004 to evaluate $\theta_E$ from \eqref{eq:enstein_ring_eqn}. Galaxy ESO 325-G004 has mass $M \approx 1.50 \times 10^{11} M_{\odot}$~\cite{Smith:2005pq,Smith:2013ena}, which contains the central black hole as well as the galaxy's luminous and dark matter.
 This galaxy serves as a lens galaxy at redshift $z_l=0.035$ lensing a background source at $z_s=2.141$.

We adopt a flat $\Lambda$CDM cosmology with Hubble constant $H_0=70.4\,\mathrm{km}\,\mathrm{s^{-1}}\,\mathrm{Mpc^{-1}}$ to determine the angular diameter distances, yielding $D_L = 154.4$ Mpc and $D_{LS} = 2866$ Mpc. The observed Einstein ring radius is:
\begin{equation}
\theta_{E, \text{obs}} = 2.85 \pm 0.40 \text{ arcsec}\,.
\end{equation}
\begin{figure}[h!]
\centering
    \begin{subfigure}{0.48\linewidth}
        \includegraphics[width=1.0\textwidth]{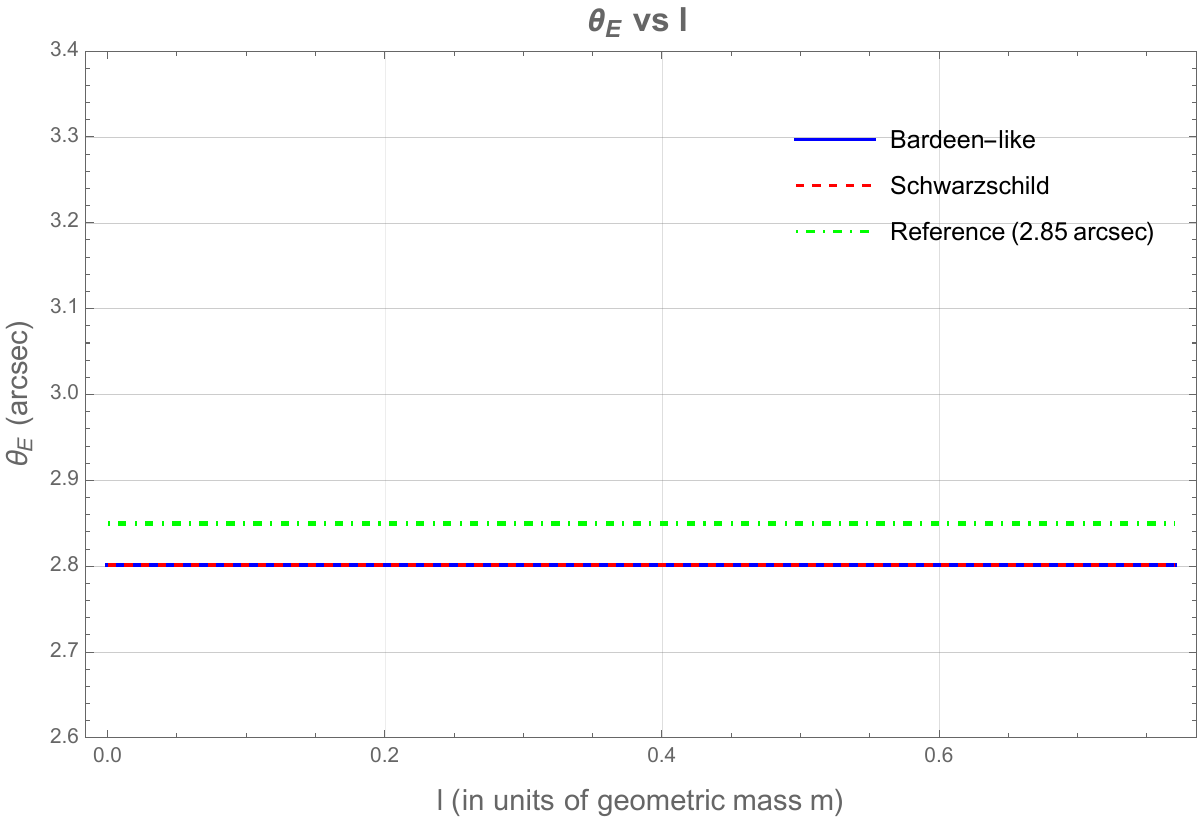}
        \caption{}
        \label{fig:theta_E_bardeen}
    \end{subfigure}
    \hfill
    \begin{subfigure}{0.48\linewidth}
        \includegraphics[width=1.0\textwidth]{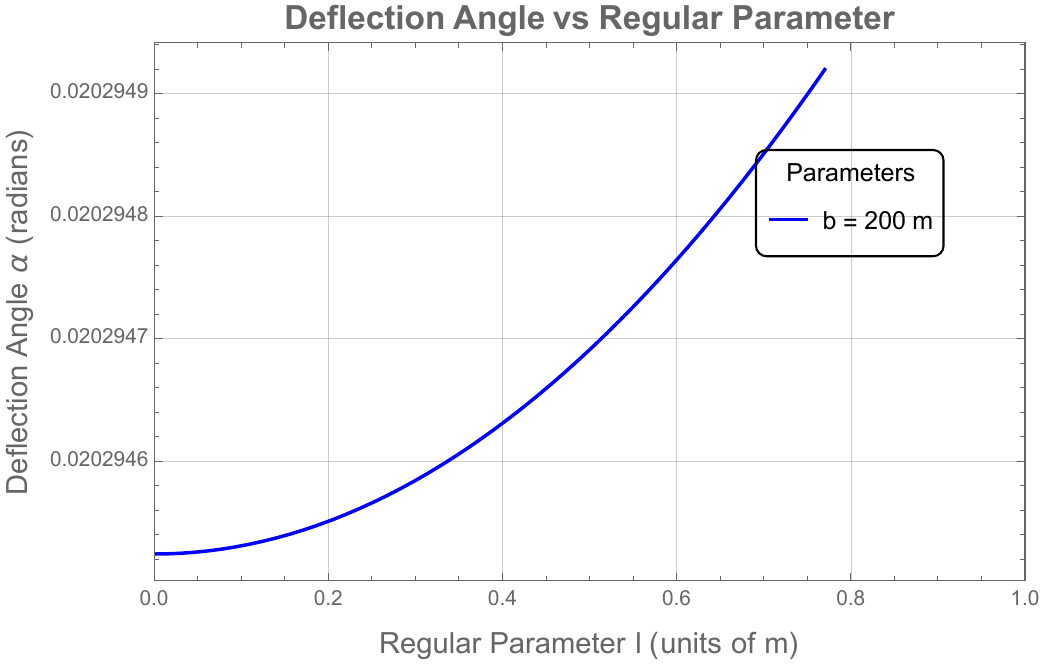}
        \caption{}
        \label{fig:deflection_impact_ell}
    \end{subfigure}
	\caption{(a) Comparison of the Einstein ring radius $\theta_E$. Blue dotted line represents the Bardeen-like black hole, while the red dotted line represents the Schwarzschild case. Both fall within the observational $1\sigma$ error bars of ESO 325-G004 (green region). (b) Deflection angle at a fixed impact parameter ($b=200m$) as a function of $\ell$, illustrating the monotonic enhancement of the deflection.}
	\label{fig:observables_weak}
\end{figure}
We solve \eqref{eq:enstein_ring_eqn} numerically to determine the predicted Einstein ring $\theta_E$ as a function of $\ell$. The results are shown in Fig.~\ref{fig:theta_E_bardeen}. The predicted radius for the Bardeen-like black hole (blue dotted line) lies well within the uncertainty over the $\ell$ range considered. It also runs nearly parallel to the Schwarzschild prediction (red dotted line). This consistency is expected because the impact parameter for galaxy-scale lensing is very large ($b \gg m$), meaning the $\ell$-dependent corrections are highly suppressed.


\bibliography{QI}{}
\bibliographystyle{aasjournalv7}



\end{document}